\documentclass[preprint,eqsecnum,aps,keywords]{revtex4}
\usepackage{dcolumn}
\usepackage{graphicx}
\usepackage{amsmath}
\usepackage{amssymb}
\usepackage{epsfig}
\usepackage{subfigure}
\usepackage{float}
\usepackage{latexsym}
\usepackage{rotating}
\begin{document}
\title{Perturbative stability of catenoidal soap films}
\author{
Soumya Jana\footnote{Electronic address: {\em soumyajana.physics@gmail.com}}${}^{}$}
\affiliation{Department of Physics \\
Indian Institute of Technology, Kharagpur 721 302, India}
\author{
Sayan Kar\footnote{Electronic address: {\em sayan@iitkgp.ac.in}}${}^{}$}
\affiliation{Department of Physics and Centre for Theoretical Studies  \\
Indian Institute of Technology, Kharagpur 721 302, India}

\begin{abstract}
\noindent The perturbative stability of catenoidal 
soap films formed between parallel, equal
radii, coaxial rings is studied using analytical and
semi-analytical methods. 
Using a theorem on the nature of eigenvalues for a class of 
Sturm--Liouville operators,
we show that for the given boundary conditions, 
azimuthally asymmetric perturbations are stable, while 
symmetric perturbations lead to an instability--a result demonstrated
in Ben Amar et. al \cite{benamar} using numerics and experiment. Further, 
we show how to obtain the lowest real eigenvalue of perturbations,
using the semi-analytical Asymptotic Iteration Method (AIM).
Conclusions using AIM support the analytically obtained result as well
as the results in \cite{benamar}.
Finally, we compute the 
eigenfunctions and show, pictorially, 
how the perturbed soap film evolves in time. 
\end{abstract}
\maketitle

\section{Introduction}
\label{sec:intro}
Soap films and bubbles have, over the years, been a topic of active interest
both in pedagogy and in research, in mathematics as well as in physics. 
In the eighteenth and nineteenth
century, it was Lagrange \cite{lagrange} and Plateau \cite{plateau} 
who pioneered the study of such {\em minimal surfaces} which 
eventually led to
the well-known {\em Plateau problem} in mathematics. Plateau, in fact was also the
first to perform a series of elegant experiments with soap films which
served as a basis for future experimental and mathematical investigations. 
Further details on the science of soap films and bubbles from a
physicists' viewpoint can
be found in the well-known book of Isenberg \cite{isenberg}. On the
mathematical front, the book by Osserman \cite{osserman}, as its title
suggests, provides a survey on  minimal surfaces. 

\noindent Our interest in this article is focused on one specific
soap film configuration--the film formed between a pair of
parallel, coaxial, equal radii rings (see Fig. 1). Geometrically, we know
that
the surface spanned by the film is a catenoid--a minimal (zero mean curvature)
surface. Following original work by Plateau, an extensive analysis on this class of films was done
in 1980 by Durand \cite{durand}. More recently, 
some theoretical and experimental work has been reported in \cite{ito}.
 
\begin{figure}[h]
 \begin{center}
 \begin{tabular}{cc}
        \resizebox{60mm}{!}{\includegraphics{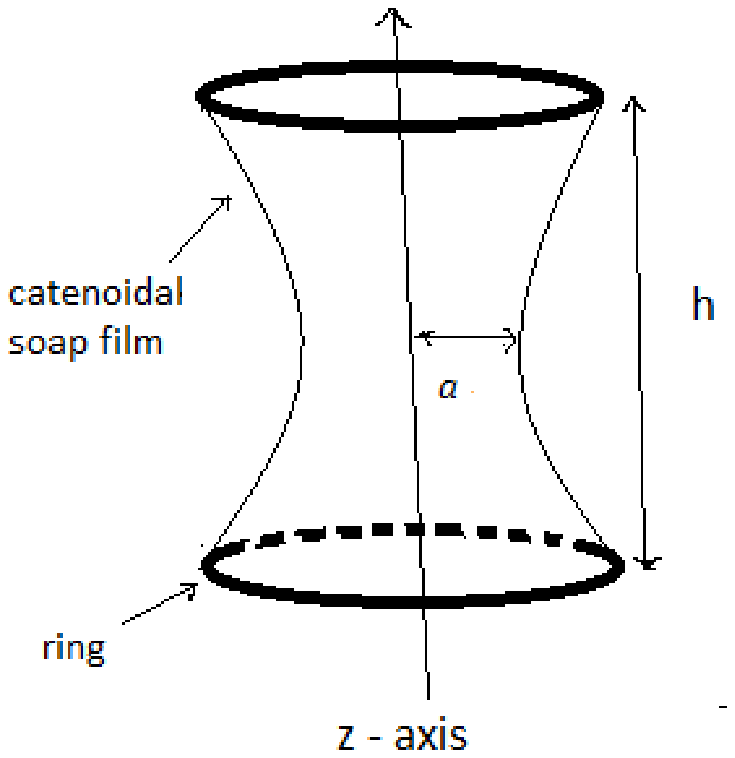}}&
        \resizebox{60mm}{!}{\includegraphics{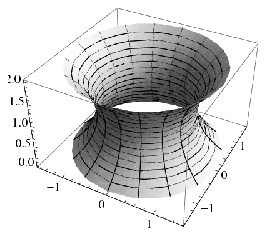}}
        
 \end{tabular}
 \caption{In the left figure, a rough sketch of a catenoidal soap film 
suspended 
between two coaxial circular rings of radius $r_0 = a\cosh(\frac{h}{2a})$, 
showing the parameters (the distance $h$ between the two rings and the minimum radius of the film $a$ ) \cite{ito}. In the right figure, 
a {\em Mathematica 7.0} 
generated catenoid (with $h= 2.0$ and $a = 1.0$) is shown.}
 \label{fig:catenoid}
 \end{center}
 \end{figure}
\noindent Finding  the shape of such a soap film is a standard problem in the
calculus of variations \cite{goldstein}. The shape is obtained 
by rotating a catenary curve about the vertical axis, to obtain a 
catenoidal surface of revolution. 
Mathematically, one first writes down the surface energy functional $V[S]$ for the 
soap film configuration, given as,
\begin{equation}
V[S]=2\pi \sigma \int^{h}_{0}r\sqrt{1+r_{z}^{2}}\,\,dz
\label{eq:1}
\end{equation}
where  $r_{z}=\frac{dr}{dz} $, $\sigma$ is the surface tension and $r(z)$ is the radius at an axial distance $z$ from the ring at $z=0$ 
(the other ring is at $z=h$). The Euler--Lagrange equation obtained from the first
variation is then solved to find the surface. 

An obvious and important question to ask is -- 
what happens if we give a small perturbation about the extremal configuration?
The 
catenoidal shape is sustained if the configuration is stable. If it is unstable,
it collapses to two disconnected planar discs, 
which is also a solution of the Euler-Lagrange equation.
To understand the stability question, which was discussed by Plateau and
later in \cite{durand}, we expand 
the surface energy functional about the extremal configuration in the following way :
\begin{equation}
V[S]= V[S_0]+\delta V[S_0]+ \delta^{2}V[S_0]+.......
\label{eq:2}
\end{equation} 
The different terms in the right hand side of the above equation are 
the zeroth, first and second variation terms. 
Setting the first variation term 
($\delta V[S_0] $) to zero  gives the Euler-Lagrange equation
which determines the extremal configuration. The next, second variation term 
$\delta^{2}V[S_0] $, is crucial for us because
it  determines the stability of the film under perturbations.

In \cite{durand} it was shown how the second variation leads to an 
associated eigenvalue problem of the Sturm-Liouville type. 
Therefore, knowing the sign of the 
lowest eigenvalue would determine the stability of such soap films. 
If the lowest eigenvalue is negative then the catenoidal configuration is 
unstable. On the other hand, a positive lowest eigenvalue
confirms its stability.
Durand \cite{durand} analyzed in detail the stability question
under azimuthally {\em symmetric} perturbations. Much later, in 1998,
Ben Amar et. al \cite{benamar} studied the stability and vibrations of
catenoid-shaped smectic films, both numerically
and experimentally.
In our work, we confirm the results in \cite{benamar} by using 
exclusively analytical and semi-analytical methods. 
We also trace the time evolution of the film and
pictorially demonstrate our conclusions on stability.

\section{Eigenvalue equation for azimuthally asymmetric perturbation} 
\label{sec:ode}
The catenoidal configuration of the soap film suspended between two parallel,
co-axial 
circular rings is described by the following equation of a catenary curve,
\begin{equation}
r(z)=a\cosh(\frac{z}{a}-\frac{h}{2a})
\label{eq:3}
\end{equation}
where, we have considered two rings of identical radius $r_0 = a\cosh(\frac{h}{2a})$.
The azimuthally asymmetric perturbed configuration is described by,
\begin{equation}
f(z,\phi) = r(z)+g(z,\phi)
\label{eq:4}
\end{equation}
where $\phi $ is the azimuthal angle in the cylindrical co-ordinate system. 
$g(z,\phi)$ is the azimuthally asymmetric perturbation with boundary 
condition: $g(0,\phi)=g(h,\phi)=0$ and $g(z,\phi)=g(z,\phi+2\pi)$.

\noindent The surface energy functional for this case comes from the following formula \cite{ito} for any general surface area (using
cylindrical coordinates and ${\bf{f}}=(f(z,\phi)\cos\phi ,f(z,\phi)\sin\phi ,z)$),
\begin{equation}
S=\int \vert\frac{\partial{\bf{f}}}{\partial z}\times\frac{\partial{\bf{f}}}{\partial \phi}\vert d\phi dz
\label{eq:5}
\end{equation}
Thus we have,
\begin{equation}
V[S]=2\sigma \int^{h}_{0}\int^{2\pi}_{0}\sqrt{f^{2}(f_{z}^{2}+1)+f^{2}_{\phi}}d\phi dz
\label{eq:6}
\end{equation} 
\noindent where, $f_{z}=\frac{\partial f}{\partial z} $ and $f_{\phi}=\frac{\partial f}{\partial \phi} $. Hence, in this case, after a Taylor expansion 
of Eq.~(\ref{eq:6}) about the extremal configuration (Eq.~(\ref{eq:3})) and 
performing the partial integration using the boundary conditions 
mentioned earlier, the second variation of the surface energy 
corresponding to the extremal surface $S_0$ becomes \cite{durand},
\begin{equation}
\delta ^2V[S_0]=\sigma\int_{0}^{2\pi}\int_{-u_0}^{u_0}\frac{g_{u}^2+g_{\phi}^{2}-g^2}{\cosh ^2 u}dud\phi
\label{eq:7}
\end{equation}
where $u(z)=\frac{z}{a}-\frac{h}{2a}$ , ($z[0,h]\longrightarrow u[-u_0,u_0] ,u_0 =\frac{h}{2a} $).
The equations~(\ref{eq:5}),(\ref{eq:6}) and (\ref{eq:7}) are also verified in detail 
using the general discussions on minimal surfaces given in \cite{beeson}.

\noindent  $g(z,\phi)$ is related to the 
infinitesimal displacement $\xi(u)$ normal to the extremal surface of the film, by the following equation given in \cite{durand} , 
\begin{equation}
g(u,\phi)=\xi(u,\phi)\cosh(u)
\label{eq:8}
\end{equation}
Eq.~(\ref{eq:8}) can be rewritten using $g(u,\phi)=g(u,\phi+2\pi)$ as,
\begin{equation}
g(u,\phi)=\zeta(u)\left(
\begin{array}{c}
\cos(m\phi)\\ \sin(m\phi) 
\end{array}\right)\cosh(u)
\label{eq:9}
\end{equation}
where $m= 0,1 ,2,3.....$

\noindent The eigenvalue problem related to the second variation (Eq.~(\ref{eq:7})),
as given in \cite{durand}, is,
\begin{equation}
L\psi_{n}\equiv-\frac{d^{2}\psi_n}{du^2}+(m^2-\frac{2}{\cosh^{2}(u)})\psi_{n} = \lambda_{n}\cosh^{2}(u)\psi_{n}
\label{eq:10} 
\end{equation} 
where, $\psi_n$, the eigen function of Sturm-Liouville operator $L$, 
is related to $\zeta$ through the following equation,
\begin{equation}
\zeta(u)=\sum^{\infty}_{n=1}c_{n}\psi_{n}
\label{eq:11}
\end{equation}
As stated before, the stability of the soap film depends on the sign of the 
lowest eigenvalue $\lambda_{1}$. 
In subsequent sections, we analyze the  Eq.~(\ref{eq:10}) 
both analytically and numerically in order to know about the nature
of the eigenvalues
and the eigenfunctions. 

\section{A result on Sturm-Liouville operators}
\label{sec:rsl}
\noindent We now state a theorem from the literature \cite{kuiper} 
on Sturm--Liouville operators which we use later to learn about the 
nature of the eigenvalues.

\noindent {\underbar{\bf{ Theorem :}}} Let Eqs.~(\ref{eq:rsl}) (see below) 
define a regular Sturm-Liouville problem, where Eq.~(\ref{subeq:rsl1}) is the eigenvalue equation, Eqs.~(\ref{subeq:rsl2}) and (\ref{subeq:rsl3}) are general boundary conditions.
\begin{subequations}
\label{eq:rsl}
\begin{equation}
 Lv := -(p(x)v')' + q(x)v = \lambda w(x)v , \hspace{0.2in}  a < x < b
 \label{subeq:rsl1}
 \end{equation} 
\begin{equation}
 B_av := A_1v(a)+A_2v'(a)+a_1 v(b) + a_2 v'(b) = 0
 \label{subeq:rsl2}
 \end{equation}
 \begin{equation}
 B_bv := b_1v(a)+b_2v'(a)+B_1 v(b) + B_2 v'(b) = 0 
 \label{subeq:rsl3}
 \end{equation}
\end{subequations}
\noindent $v'=\frac{dv}{dx} $. 
It is assumed that the interval $[a,b]$ is bounded 
(i.e. $-\infty<a<b<\infty $), the boundary conditions are linearly independent, all coefficients are real and the functions $p, p', q, w$ are continuous 
in the domain $[a,b]$. Also, $p(x)>0$ and $w(x)>0$ in $[a,b]$. 
If the boundary conditions are {\em separated} (i.e. $a_1 = a_2 = b_1 = b_2 = 0 $) and $A_1A_2 \leq 0 $ and $B_1B_2 \geq 0$, then, using the Green's identities \cite{kuiper,dennery} for the Sturm-Liouville operator, we can show that for 
any eigenpair $(\lambda, v)$,
\begin{equation}
 \lambda \geq \frac{min[q(x): a \leq x \leq b]}{max[w(x):a \leq x \leq b ]}
 \label{eq:theom}
\end{equation}
For a proof of the above theorem see Appendix~\ref{app:theorem}.

\noindent {\underbar {\bf{Application:}}} \\In our problem, $x = u$, 
$p(u) = 1 $, $q(u)= - \frac{2}{\cosh^{2}(u)} + m^2 $, $w(u) = \cosh^{2}(u) $, 
$a = -u_0$, $b = u_0 $ and boundary conditions are separated, with $A_2 = B_2 = 0$ and $A_1 , B_1 \neq 0$. 
Therefore, our problem is a regular Sturm-Liouville problem satisfying the 
properties mentioned in the theorem quoted above.
Thus, for $\lambda_1$ to be a negative eigenvalue, we need
 \begin{equation}
 min[q(u):  -u_0 \leq u \leq u_0] < 0  \Rightarrow  \cosh^2(u_0) < \frac{2}{m^2} 
 \label{eq:stabilitycondition}
 \end{equation}
For $m=0 $, the above condition is always satisfied. 
If $m=1 $, we have, $1<\cosh^{2}(u_0)<2 $ . However, for the existence of a 
negative eigenvalue, $u_0$ must be greater than $1.2$, since this 
is the critical value of $u_0$ for which the 
lowest eigenvalue $\lambda_{1}$ is zero, for $m=0$. Only above this value of 
$u_0$ does $\lambda_{1}$ become negative for $m=0$ \cite{durand}. 
Note that $\cosh^{2}(1.2)=3.278$. Thus, for $m=1$, a negative eigenvalue 
does not exist. Further, for $m\geq 2$, the inequality (\ref{eq:stabilitycondition}) becomes, $\cosh^2(u_0)<\frac{1}{2}$ which is impossible because
$\cosh^2(u_0)\geq 1$ always. Therefore for all $m\geq2$, 
no negative eigenvalues exist.

\noindent The analysis above leads us to the result that, only the 
$m = 0$ mode can be unstable.

\section{Semi-analytical, numerical analysis}
\label{sec:numerical}
\subsection{Eigenvalues using AIM}
\noindent The eigenvalues of the above Sturm-Liouville problem can be 
computed using the Asymptotic Iteration Method (AIM) for eigenvalue 
problems \cite{ciftci}, which is briefly discussed in Appendix~\ref{app:aim}. 
In our case the differential equation is,
\begin{equation}
\frac{d^2 \psi _n }{du^2} + (\lambda _n \cosh ^2 u -m^2+\frac{2}{\cosh ^2 u})\psi _n =0
\label{eq:17}
\end{equation}
Since $\psi _n (-u_0)$= $\psi _n (u_0)=0 $, $\psi _n(u)$ includes two factors 
which are $(u+u_0)$ and $(u-u_0)$, the following transformation can be 
used for the present case,
\begin{equation}
\psi _n (u)= (u^2-u_0^2)\varphi _n(u)
\label{eq:18}
\end{equation}
Using the above transformation in Eq.~(\ref{eq:17}), we obtain the differential 
equation satisfied by $\varphi _n (u)$,
\begin{equation}
\frac{d^2\varphi _n}{du^2}=-\frac{4u}{u^2-u_0^2} \frac{d\varphi _n}{du}-(\frac{2}{u^2-u_0^2}+\lambda _n\cosh ^2u-m^2 +\frac{2}{\cosh ^2u})\varphi _n
\label{eq:19}
\end{equation}
Comparing Eq.~(\ref{eq:19}) with Eq.~(\ref{eq:12}),using Eqs.~(\ref{eq:14}), (\ref{eq:15}) and (\ref{eq:16}), 
we can calculate $\delta$ (see Appendix B for definition) 
and using a particular chosen value $ u=u_{ch} $ between $-u_0$ to $u_0$,  
$\delta$ becomes a function of $\lambda _n$ and finally, $\lambda _n$ can be 
computed by finding the roots of $\delta =0$. 
In this case, 
while computing the eigenvalues $u_{ch}=0$ is used. In this way, 
eigenvalues were computed for different $u_0$.

\subsection{Analytical check}
The eigenvalue equation related to this problem can be solved analytically 
for two limiting cases of the $m=0$ mode. For this azimuthally symmetric case 
($m=0$), 
when $u_0 \rightarrow u_c = 1.2$ it is shown in \cite{durand} that,
\begin{subequations}
\label{eq:analytic1}
\begin{equation}
\lambda_{1} \approx 3.598 (\frac{h_c-h}{2r_0})^{\frac{1}{2}},\hspace{0.2in} u_0 < u_c 
\label{eq:22}
\end{equation}
\begin{equation}
\lambda_{1} \approx - 3.598 (\frac{h_c-h}{2r_0})^{\frac{1}{2}},\hspace{0.2in} u_0 > u_c
\label{eq:23} 
\end{equation}
\end{subequations}
\noindent Again if $u_{0} \rightarrow 0$ then,
\begin{equation}
\lambda_{1}\rightarrow(\frac{\pi}{2u_0})^2-2
\label{eq:24}
\end{equation}
\begin{table*}
\caption{\label{tab:table1}Comparison between the lowest eigenvalues for $m=0$ 
mode obtained from {\em Asymptotic Iteration Method} using 16 iterations in {\em Mathematica}, with the analytical values from Eqs.~(\ref{eq:analytic1})and (\ref{eq:24})}
\begin{ruledtabular}
\begin{tabular}{ccccccc}
 \multicolumn{3}{c}{\textrm{$u_{0}\rightarrow u_c$}}&&\multicolumn{3}{c}{\textrm{$u_{0}\rightarrow 0$}}\\
 \textrm{$u_0$}&\textrm{$\lambda_{1}$ (Eq.~\ref{eq:analytic1})}&\textrm{$\lambda_{1}$ (AIM)}
 &&\textrm{$u_0$}&\textrm{$\lambda_{1}$ (Eq.~\ref{eq:24})}&\textrm{$\lambda_{1}$ (AIM)}\\ \hline
 1.1 & 0.212 & 0.244 && 0.01 & 24672.01  & 24671.70\\
 1.15 & 0.104  & 0.107 && 0.02  & 6166.50 & 6166.18\\
 1.2 & 0.0  & -0.01 && 0.03  & 2739.56 & 2739.23 \\
 1.25 & -0.103 & -0.114 && 0.04  & 1540.12 & 1539.80 \\
 1.30 & -0.202 &  -0.210 && 0.05 & 984.96  & 984.64\\
 \end{tabular}
\end{ruledtabular}
\end{table*}
The above check shown in Table~\ref{tab:table1} 
confirms that we can indeed use AIM to compute the eigenvalues for
(Eq.~(\ref{eq:17})).
 
 \subsection{Convergence of results in AIM}

\noindent On performing the numerical calculations using AIM, 
we found that only 10 iterations were possible in {\em Mathematica 7.0} in a 32-bit system. However, 
using the
{\em Improved Asymptotic Iteration Method} (IAIM) 
(see Appendix~\ref{app:iaim}), 17 iterations could be carried out
for the problem at hand. We noted that for higher value of 
$u_0$, the results using 10 iterations are different from the results obtained 
from 16 iterations. This deviation increases with
increasing $u_0$. The reason is that 10 iterations are not 
sufficient 
to guarantee a convergence towards the result. We have inspected the 
convergence of our result for a number of iterations choosing $u_0 = 1.1$.
We note that the difference in results between 8th 
and 10th iteration is 0.06374, between 10th and 12th iteration it 
is 0.031732 , between 12th and 14th iteration it is 0.015443 and  
between 14th and 16th it is 0.007601. Thus, as the iteration number 
increases, the result converges more and more. After each iteration, $\eta$ or $s$ (see Appendix~\ref{app:aim} for definition) become zero for $u_{ch}=0$ 
in an alternating fashion, resulting in the same lowest eigen value for 
two consecutive iterations. Therefore,  we compare the $n$th and the
($n+2$)th iterations. After 16 iterations, the result 
is seen to converge up to the second decimal point. 
If we wish to have
even better convergence, we need to further increase the number of iterations.

\noindent However, let us inspect the same facts for $u_0 = 1.5$ 
(Table~\ref{tab:table2}).
\begin{table*}
\caption{\label{tab:table2}Result of different number of iterations for
 finding the lowest real eigenvalue $\lambda_1$ for $u_0=1.5 , m=0$.}
\begin{ruledtabular}
\begin{tabular}{ccccc}
 \textrm{number of iterations}&\textrm{$\lambda_1$}& &\textrm{number of iterations}&\textrm{$\lambda_1$} \\ \hline
 8 & -0.756028 & & 13 & -0.704753\\ 
 9 &  -0.756028 & & 14 & 0.177329\\ 
 10 & 0.448441 & & 15 &  0.177329\\ 
 11 & 0.448441 & & 16 & -0.660185\\ 
 12 & -0.704753 & & 17 & -0.660185\\ 
 \end{tabular}
\end{ruledtabular}
\end{table*}
In this case the difference in results between 9th and 10th iteration 
is 1.204469, between 10th and 12th iteration is 1.153194, 
between 13th, 14th iteration it is 0.882082 and 
between  15th and 16th iteration it is 0.837514.
We note that the result is indeed converging as the number 
of iterations increase. But, unlike the previous case
(i.e. for $u_0 = 1.1$), where 
after 16 iterations, the result converged upto the second decimal point, 
we find that for $u_0= 1.5$ the result does not converge even to 
the first decimal point! In addition, the difference between the results 
of 15th 
and 16th iteration is of the order of the actual eigenvalue. 
Clearly, for this case, we need a larger number of iterations 
to get the correct eigenvalue. 
\subsection{A way to overcome the limitation of AIM}  
\noindent In the previous section, we have seen that for $u_0 > 1.2$ we 
need many more iterations than 16. However, using Mathematica we were unable 
to go beyond 17 iterations. In order to overcome this limitation, 
we have employed a heuristic approach which is
elaborated in the Appendix~\ref{app:heuristic}.

\begin{figure}[h]
\begin{center}
\mbox{\epsfig{file=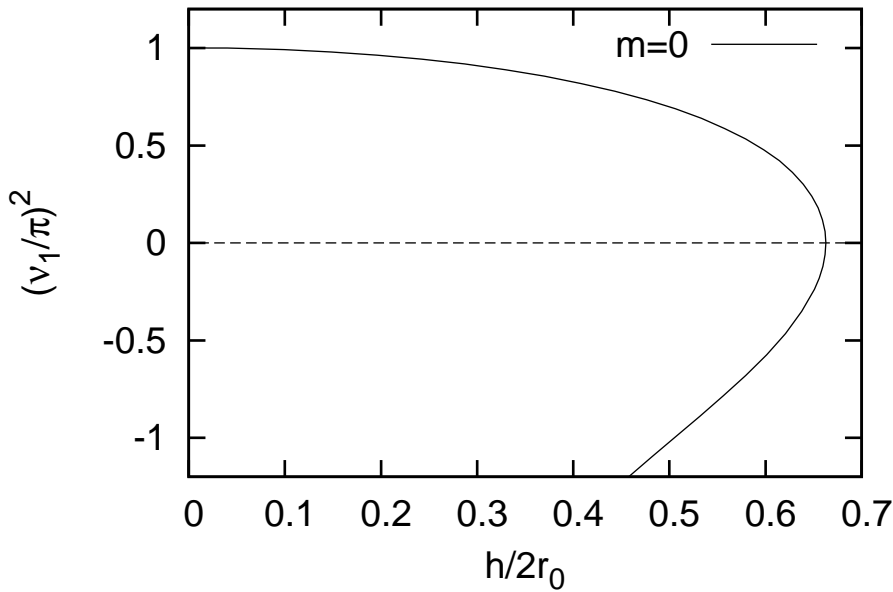,width=5.in,angle=360}}
\end{center}
\caption{Plot of lowest eigenvalue(corrected) in terms of square of reduced frequency and as a function of $\frac{h}{2r_0}$ }
\label{fig:correct eigenvalue}
\end{figure}

\noindent Using the combination of AIM and the above-stated method,
 we finally obtain the correct eigenvalues for $m=0$. 
 We have plotted $(\frac{\nu_1}{\pi})^2$ as a function of $\frac{h}{2r_0}$ in Fig.~\ref{fig:correct eigenvalue} (where , $\nu_1$ is the reduced frequency 
defined in \cite{durand}). Fig.~\ref{fig:correct eigenvalue} exactly matches with the plot shown in \cite{durand} (see Fig. 4 there). This ensures that our method 
is reliable and can be used to find the eigenvalues and analyse the 
stability of soap films between the two rings.

\section{Result of numerical analysis}
\label{sec:result}
\begin{figure}[h]
\begin{center}
\mbox{\epsfig{file=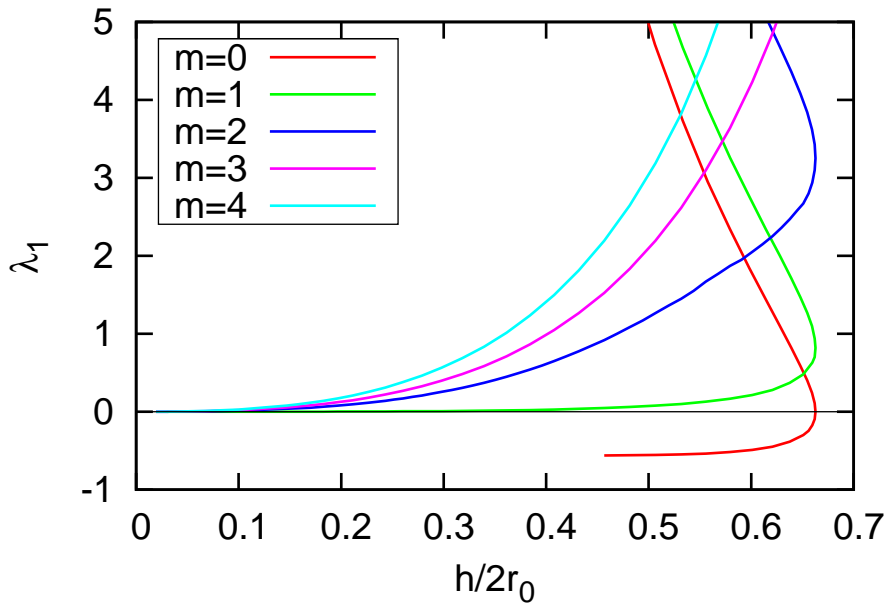,width=5.in,angle=360}}
\end{center}
\caption{Plot of lowest eigenvalue $\lambda_1$ as a function of $\frac{h}{2r_0}$ for different $m$}
\label{fig:allmodes}
\end{figure}

In this section we discuss the plots for the lowest eigenvalues as a function 
$\frac{h}{2r_0}$ for different azimuthally asymmetric modes ($m\neq 0$), 
where we have shown the behaviour for large $u_0$. 
The plots in Fig.~\ref{fig:allmodes} show that only the 
$m=0$ plot goes below the zero eigenvalue line (it crosses the zero eigenvalue line at $u_0 = 1.2$ , alternatively $\frac{h}{2r_0}=0.663$). All other ($m=1,2,3,4$) $m\neq 0$ curves asymptotically approach the zero value, 
for $u_0 \rightarrow \infty$ or $\frac{h}{2r_0}\rightarrow 0$. Thus,
the result of numerical analysis agrees with the fact that only the 
$m=0$ mode can 
be unstable, a fact which we obtained analytically as well. This is the
central result in our article.

\section{Perturbed configurations}
\label{sec:perfig}
Let us now try and see if we can obtain the perturbed configurations
and visualize their time evolution.
If an infinitesimal displacement is given to the  extremal configuration of 
the soap film which is  described by Eq.~\ref{eq:3}, then the film will have 
a small motion about it. For the azimuthally symmetric case ($m=0$) the 
general expression for the displacement about the extremal configuration, which describes this small 
motion of the soap film, is (given in \cite{durand}),
 \begin{equation}
 \xi(u,t)=\sum^{\infty}_{n=1}(a_n\cos(w_{n}t)+\frac{b_n}{w_n}\sin(w_{n}t))\psi_{n}(u)
 \label{eq:25}
 \end{equation}
 where
 \begin{equation}
 a_n=\int^{u_0}_{-u_0}\psi_{n}(u)\xi(u,0)\cosh^{2}udu
 \label{eq:26}
 \end{equation}
 \begin{equation}
 b_n=\int^{u_0}_{-u_0}\psi_{n}(u)\xi_{t}(u,0)\cosh^{2}udu
 \label{eq:27}
 \end{equation}
 \begin{equation}
 w_n=(\frac{2\sigma}{\rho a^2}\lambda_{n})^{\frac{1}{2}}=(\frac{2\sigma}{\rho h^2})^{\frac{1}{2}}\nu_{n}
 \label{eq:28}
 \end{equation} and $\xi_{t}=\frac{\partial \xi}{\partial t}$.
The frequency of oscillation $w_n$ is related to the 
dimensionless reduced frequency $\nu_n$ by Eq.~(\ref{eq:28}), 
where $\rho$ is the mass of the film per unit area. 
If $u_0 < u_c $ ($u_0 = \frac{h}{2a}$ and $u_c =\frac{h_c}{2a}=1.2$ corresponds to $\lambda_1 = 0$), then $w_{1}^{2} > 0$, consequently $w_1$ is real and 
the motion described by Eq.~(\ref{eq:25}) is the small oscillation of the 
soap film about its stable configuration.

If $u_0 > u_c$ ,$w_{1}^{2} < 0$ ,$w_1$ is pure imaginary and the first term in Eq.~(\ref{eq:25}) is hyperbolic,
\begin{equation}
 \xi(u,t)=(a_{1}\cosh |w_1|t +\frac{ b_{1}}{w_1}\sinh |w_1|t )\psi_{1}(u) + \sum^{\infty}_{n=2}(a_n\cos(w_{n}t)+\frac{b_n}{w_n}\sin(w_{n}t))\psi_{n}(u)
 \label{eq:29}
\end{equation}
Thus, the first term grows in time and the configuration is unstable--
it collapses into a configuration consisting of two plane discs.

Now if we consider an azimuthally asymmetric displacement,
the expression for the displacement may be written as:
\begin{equation}
\xi(u,\phi ,t) = \sum^{\infty}_{m=0}\sum^{\infty}_{n=1} (a_{mn}\cos (w_{mn}t+m\phi)+b_{mn}\sin (w_{mn}t+m\phi))\psi_{mn}(u)
\label{eq:30}
\end{equation} 
where
\begin{equation}
a_{0n} = \frac{1}{2\pi}\int^{u_0}_{-u_0}\int^{2\pi}_{0}\xi(u,\phi ,0)\psi_{0n}(u)\cosh ^{2}u \,\,d\phi \,\,du
\label{eq:31}
\end{equation} 
\begin{equation}
b_{0n} = \frac{1}{2\pi w_{n0}}\int^{u_0}_{-u_0}\int^{2\pi}_{0}\xi_{t}(u,\phi ,0)\psi_{0n}(u)\cosh ^{2}u\,\,d\phi\,\, du
\label{eq:32}
\end{equation} 
\begin{equation}
a_{mn} = \frac{1}{\pi}\int^{u_0}_{-u_0}\int^{2\pi}_{0}\xi(u,\phi ,0)\psi_{mn}(u)\cosh ^{2}u\cos (m\phi)\,\, d\phi\,\, du
\label{eq:33}
\end{equation} 
\begin{equation}
b_{mn} = \frac{1}{\pi}\int^{u_0}_{-u_0}\int^{2\pi}_{0}\xi(u,\phi ,0)\psi_{mn}(u)\cosh ^{2}u\sin (m\phi)\,\,d\phi\,\,du
\label{eq:34}
\end{equation} 
In Eq.~(\ref{eq:30}), the instability occurs only in the leading order term.

\noindent  Using the eigenvalues obtained earlier, 
we solve Eq.~(\ref{eq:10}) numerically and then plot the perturbed 
configurations. While plotting, we treat the eigenfunction itself
as the perturbation, though in a real situation the perturbation is the linear 
combination of such eigenfunctions satisfying the same boundary condition. 

\begin{figure}[h]
 \begin{center}
 \begin{tabular}{cc}
        \resizebox{60mm}{!}{\includegraphics{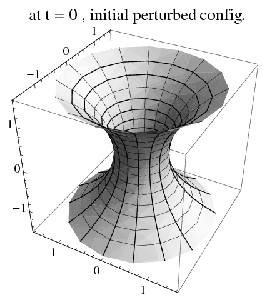}}&
        \resizebox{60mm}{!}{\includegraphics{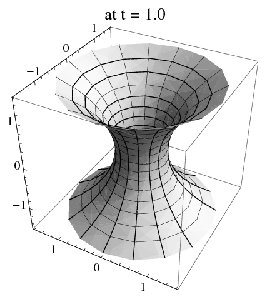}}\\
        \resizebox{60mm}{!}{\includegraphics{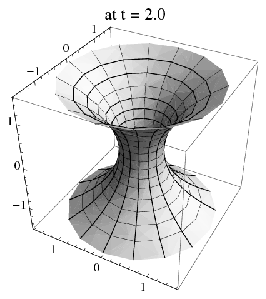}}&
        \resizebox{60mm}{!}{\includegraphics{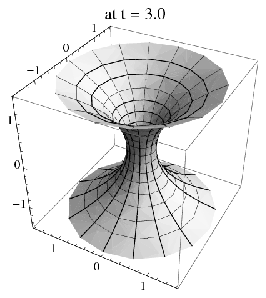}}\\
        \resizebox{60mm}{!}{\includegraphics{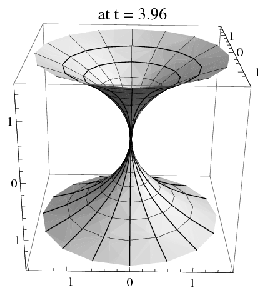}}&
        \resizebox{60mm}{!}{\includegraphics{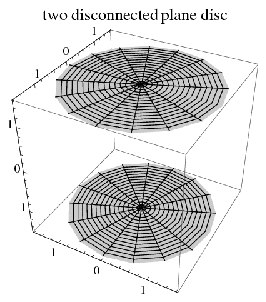}}\\
 \end{tabular}
 \caption{Unstable configuration ;$u_0 = 1.5$ ,$m=0$ , $\lambda_1 =-0.385  , \chi(-u_0)=-0.1$ ,unit of time $\frac{1}{\pi}\left( \frac{\rho h^2}{2\sigma}\right)^{\frac{1}{2}}  $  }
 \label{fig:unstable}
 \end{center}
 \end{figure}

\begin{figure}[h]
 \begin{center}
 \begin{tabular}{ccc}
        \resizebox{40mm}{!}{\includegraphics{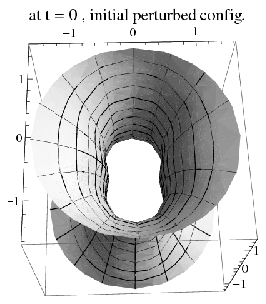}}&
        \resizebox{40mm}{!}{\includegraphics{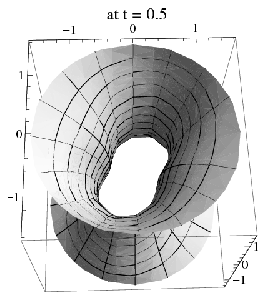}}&
        \resizebox{40mm}{!}{\includegraphics{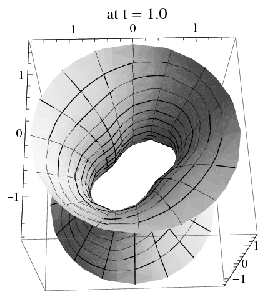}}\\
        \resizebox{40mm}{!}{\includegraphics{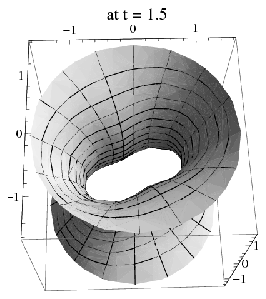}}&
        \resizebox{40mm}{!}{\includegraphics{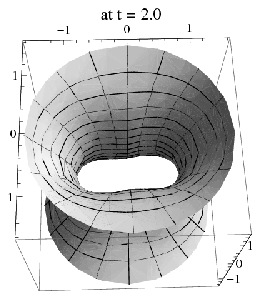}}&
        \resizebox{40mm}{!}{\includegraphics{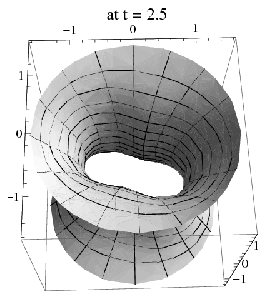}}\\
        \resizebox{40mm}{!}{\includegraphics{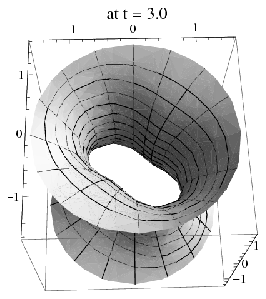}}&
        \resizebox{40mm}{!}{\includegraphics{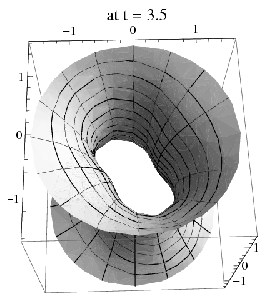}}&
        \resizebox{40mm}{!}{\includegraphics{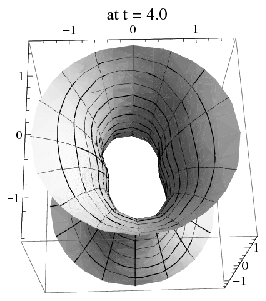}}\\
 \end{tabular}
 \caption{Stable configuration (top view) under $m=2$ mode of vibration ; $u_0 = 1.5$ ,$\lambda_1 = 2.47 , \chi(-u_0)=-0.4$, unit of time $\frac{1}{\pi}\left( \frac{\rho h^2}{2\sigma}\right)^{\frac{1}{2}}$ }
 \label{fig:stable}
 \end{center}
 \end{figure}

\begin{figure}[h]
 \begin{center}
 \begin{tabular}{cc}
        \resizebox{70mm}{!}{\includegraphics{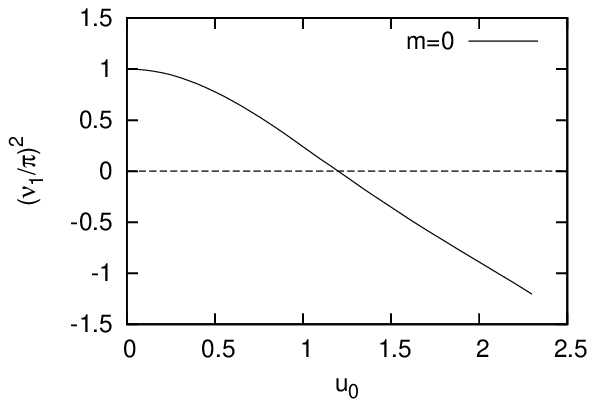}}&
        \resizebox{70mm}{!}{\includegraphics{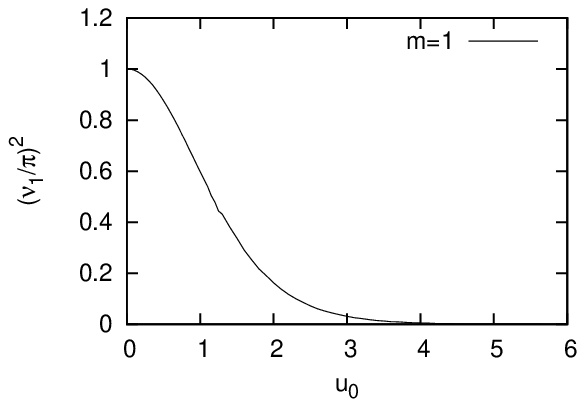}}\\
        \resizebox{70mm}{!}{\includegraphics{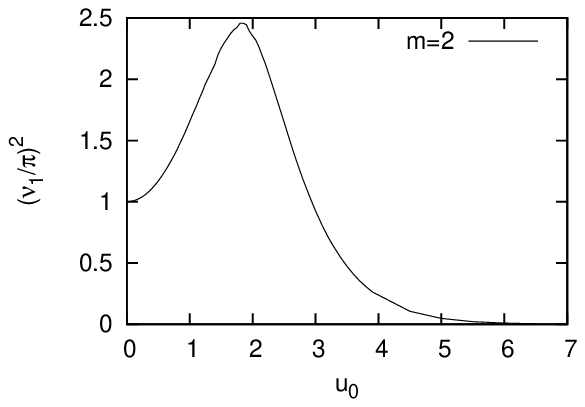}}&
        \resizebox{70mm}{!}{\includegraphics{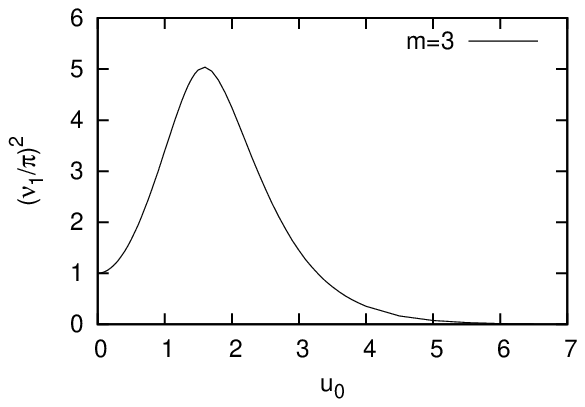}}\\
        
 \end{tabular}
 \caption{Variation of square of reduced frequency $\nu_1$ of oscillation with $u_0$ for different modes }
 \label{fig:frequency}
 \end{center}
 \end{figure}

\noindent Eq.~(\ref{eq:10}), which is a second order differential equation can be written  in terms of two first order differential equations in the following way:
\begin{equation}
\frac{d\zeta}{du}= \chi
\label{eq:35}
\end{equation} 
\begin{equation}
\frac{d\chi}{du}= -(\lambda\cosh^2(u)-m^2+\frac{2}{\cosh^2(u)})\zeta
\label{eq:36}
\end{equation} where, $\zeta\equiv\psi(u)$.
We solve the above two ordinary differential equations numerically using 
Mathematica with the initial conditions:
$\zeta(-u_0)=0$ and $\chi(-u_0)$(of arbitrary choice). Using the computed eigenfunction we evaluate the $g(u,\phi,t)$ 
via the following relation:

\noindent for the oscillating mode,
\begin{equation}
g(u,\phi,t)=\zeta(u)\cosh(u)cos(m\phi +w_{mn}t)
\label{eq:37} 
\end{equation}
for the collapsing mode,
\begin{equation}   
g(u,\phi,t)=\zeta(u)\cosh(u)\cosh(|w_{01}|t)
\label{eq:38}
\end{equation}
Finally, we plot the time evolution of the perturbed configuration using 
the following relation:
\begin{equation}
f(u,\phi,t)= r(u)+g(u,\phi,t)
\label{eq:39}
\end{equation}
In Fig.~\ref{fig:unstable} and Fig.~\ref{fig:stable} we have shown the 
collapsing mode  and oscillating mode respectively, where snapshots in time 
appear in each frame. The elapsed time mentioned in the figures are in 
units of $\frac{1}{\pi}\left( \frac{\rho h^2}{2\sigma}\right)^{\frac{1}{2}} $.  
In Fig.~\ref{fig:frequency}, we have shown how the frequency of oscillation 
varies with the boundary ($u_0$) and for different modes ($m$).
 
It is clear from the time evolution that the analytical and numerical
conclusions on stability found and stated earlier are once again 
confirmed.

\section{Summary}
\label{sec:summary} 
We have shown using analytical and semi-analytical methods,
that only the azimuthally 
symmetric($m=0$) perturbation of the catenoidal soap film 
can be unstable and this is true when 
$u_{0}=\frac{h}{2a}$ is greater than $1.2$. Physically this 
means that for any given small $m=0$ perturbation, the soap film can 
collapse in an azimuthally symmetric way into two disconnected plane discs.
On the other hand  for $m\neq 0$, the film remains stable as long as
the perturbation does not become too large. This rather counter-intuitive
result was shown first, using numerics and experiment in \cite{benamar}. Our
results re-confirm the result in \cite{benamar} using different theoretical
tools such as (a) a purely analytical method and  (b) the semi-analytical
AIM.  

\noindent One can also analyze a more general situation, 
where the radius of the rings 
are not same, using the methods employed here. 
For such a case, we have to 
alter the transformations in Eq.~(\ref{eq:18}) accordingly, 
by invoking appropriate boundary conditions. 

\noindent In addition to the above results, our work also provides an example
of the use of AIM and improved AIM in tackling eigenvalue problems
which are difficult to solve, analytically.

\section*{Acknowledgments}
SJ acknowledges the Department of Physics, IIT Kharagpur, India for 
providing him the opportunity to work on this project. 
\appendix
\section{{\em Proof} of the theorem referred in Section~\ref{sec:rsl}}
\label{app:theorem}
\noindent For a regular Sturm-Liouville problem defined by Eq.~(\ref{eq:rsl}),
\begin{equation}
\int^{b}_{a}vLudx=-vpu'\vert^{b}_{a}+\int^{b}_{a}(pv'u'+qvu)dx
\label{eq:greensid}
\end{equation}
where, $u(x),v(x)$ are eigenfunctions of the operator $L$, both satisfying the boundary conditions, Eq.~(\ref{subeq:rsl2}) and Eq.~(\ref{subeq:rsl3}). 
Since $p(x)>0$ on $[a,b] $, using $\bar{v}$,the complex conjugate of $v$, 
in place of $u$ in Eq.~(\ref{eq:greensid}) we get,  
\begin{equation}
\int^{b}_{a}vL\bar{v}'\geqslant p(a)v(a)\bar{v}'(a)-p(b)v(b)\bar{v}'(b)+\int^{b}_{a}q\mid{v}\mid^2 dx 
\label{eq:greensid1}
\end{equation}
If the boundary conditions are {\em separated}, which mean,
\begin{subequations}
\label{eq:seprtedbc}
\begin{equation}
A_{1}v(a)+A_{2}v'(a)=0
\label{subeq:seprtedbc1}
\end{equation}
\begin{equation}
B_{1}v(b)+B_{2}v'(b)=0
\label{subeq:seprtedbc2}
\end{equation}
\end{subequations}
then, $A_{1}A_{2}\leqslant 0$ implies $v(a)\bar{v}'(a)\geqslant 0$ and $B_{1}B_{2}\geqslant 0 $ implies $v(b)\bar{v}'(b)\leqslant 0$. So, the 
first two terms on the right hand side of the inequality~(\ref{eq:greensid1}) are positive. Thus, we have, 
\begin{equation}
\int^{b}_{a}vL\bar{v}'dx\geqslant \int^{b}_{a}q\vert v\vert^{2}dx
\label{eq:theormineq}
\end{equation}  
For {\em separated boundary conditions}, $L$ is {\em self adjoint} operator and hence, $v$ and $\bar{v}$ have a common real eigenvalue, say $\lambda$. 
Then, inequality~(\ref{eq:theormineq}) becomes,
\begin{equation}
\lambda \int^{b}_{a}\vert v\vert^{2}wdx\geqslant \int^{b}_{a}q\vert v\vert^{2}dx
\label{eq:theormineq1}
\end{equation}
Thus, we have the theorem given by Eq.~(\ref{eq:theom}),
\begin{equation}
 \lambda \geq \frac{min[q(x): a \leq x \leq b]}{max[w(x):a \leq x \leq b ]}
\end{equation}  
\section{\em Asymptotic Iteration Method (AIM)}
\label{app:aim}
\noindent Consider the homogeneous, linear, second order differential 
equation
\begin{equation}
y''=\eta _0 (x)y' + s_0 (x)y 
\label{eq:12}
\end{equation}
According to the Asymptotic Iteration Method, for sufficiently large 
value of $n$ ($n$ is an integer),  
\begin{equation}
\frac{s_n}{\eta _n}= \frac{s_{n-1}}{\eta_{n-1}} \equiv \alpha
\label{eq:13}
\end{equation}
where
\begin{equation}
\eta _k = \eta '_{k-1} +s_{k-1} + \eta_0 \eta _{k-1}
\label{eq:14}
\end{equation}
and
\begin{equation}
s _k = s'_{k-1} + s_ {0} \eta _{k-1}
\label{eq:15}
\end{equation}
for $k = 1,2,3,.....n$. We also define a function $\delta$ such that 
\begin{equation}
 \delta = s_n \eta _{n-1}-s_{n-1} \eta _n
 \label{eq:16}
\end{equation}
The eigenvalues can be computed by means of $\delta= 0$.
\label{app:iaim}

\noindent Due to the presence of differentiation in Eqs.~(\ref{eq:14}) 
and (\ref{eq:15}), AIM may slow down the computer and consequently 
less number of iterations 
can be performed. However, in some problems 
a large number of iterations are required
for convergence of the result. Improved Asymptotic Iteration Method, introduced in \cite{cho }, can speed up the process quite a bit. 
$\eta_{n}(x) $ and $s_{n}(x)$ can be expanded in series around $x_0$, 
which is used in Eq.~(\ref{eq:16}) to compute the eigenvalue.
\begin{subequations}
\label{eq:iaim}
\begin{equation}
\eta_{n}(x)=\sum^{\infty}_{i=0}c^{i}_{n}(x-x_0)^i
\label{subeq:iaim1a}
\end{equation}
\begin{equation}
s_{n}(x_0)=\sum^{\infty}_{i=0}d^{i}_{n}(x-x_0)^i
\label{subeq:iaim1b}
\end{equation}
\end{subequations}
where $c^{i}_{n}$ and $d^{i}_{n} $'s are the 
Taylor's series expansion coefficients. Using Eqs.~(\ref{eq:iaim}) in Eqs.~(\ref{eq:14}) and (\ref{eq:15}) we get following recursion relations (Eq.~(\ref{eq:taylorcoeff})).
\begin{subequations} 
\label{eq:taylorcoeff}
\begin{equation}
c^{i}_{n}=(i+1)c^{i+1}_{n-1}+d^{i}_{n-1}+\sum^{i}_{k=0}c^{k}_{0}c^{i-k}_{n-1}
\label{subeq:taylorc}
\end{equation}
\begin{equation}
d^{i}_{n}=(i+1)d^{i+1}_{n-1}+\sum^{i}_{k=0}d^{k}_{0}c^{i-k}_{n-1}
\label{subeq:taylord}
\end{equation}
\end{subequations}

\begin{equation}
d^{0}_{n}c^{0}_{n-1}-d^{0}_{n-1}c^{0}_{n}=0
\label{eq:iaimeigen}
\end{equation}
The equation $\delta=0$, now turns out to be Eq.~(\ref{eq:iaimeigen}), 
which can be used for the computation of eigenvalues.
 
\section{Heuristic approach for finding eigen value}
We have used a heuristic approach to find the eigenvalues in the
absence of the possibility of performing a large number of iterations.
This approach is briefly outlined below.
\label{app:heuristic}

$\bullet$ We solve the Eq.~\ref{eq:17} and plot the eigenfunction 
corresponding to the lowest eigenvalue, using the eigenvalue obtained 
from AIM. The method used for this has been discussed in Section~\ref{sec:perfig}.

$\bullet$ We use the fact that the eigenfunction must satisfy the boundary 
condition $\psi_1(-u_0)=\psi_1(u_0)=0$ and the eigenfunction will be 
symmetric.  

$\bullet$ If the plotted  eigenfunction violates the above facts 
then we change the eigenvalue slightly around the value obtained 
from AIM and go through the same procedure until we get the correct 
eigenvalue.

\begin{figure}[ht]
\centering
\subfigure[$\lambda_{1}=-0.66$ {\bf (AIM)}]
{\includegraphics[scale=0.4]{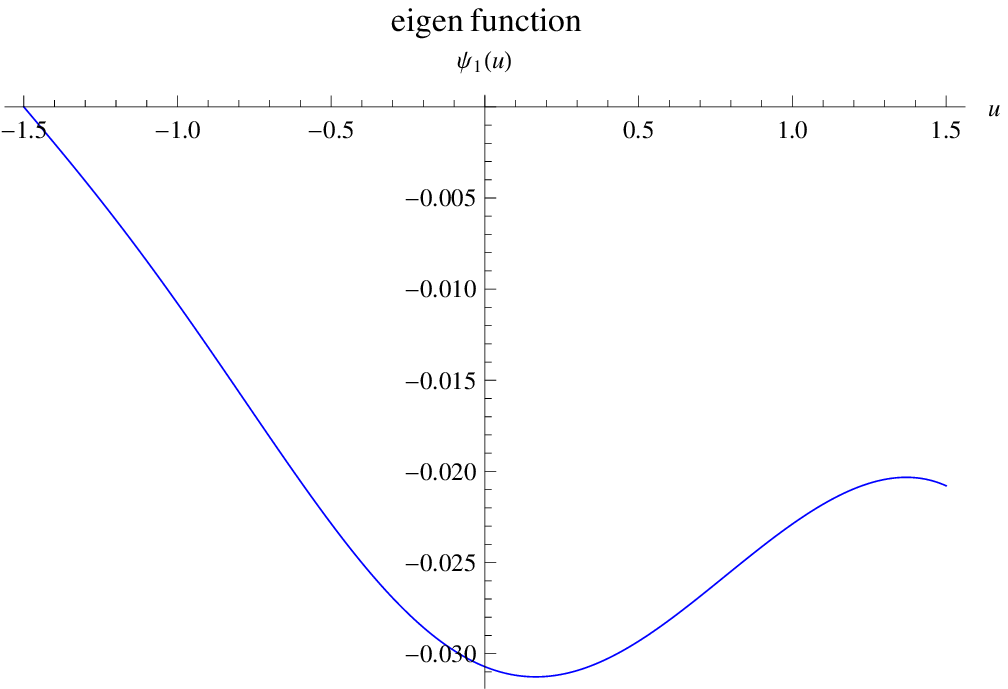}
\label{fig:subfig1}}
\subfigure[$\lambda_{1}=-0.8$]
{\includegraphics[scale=0.4]{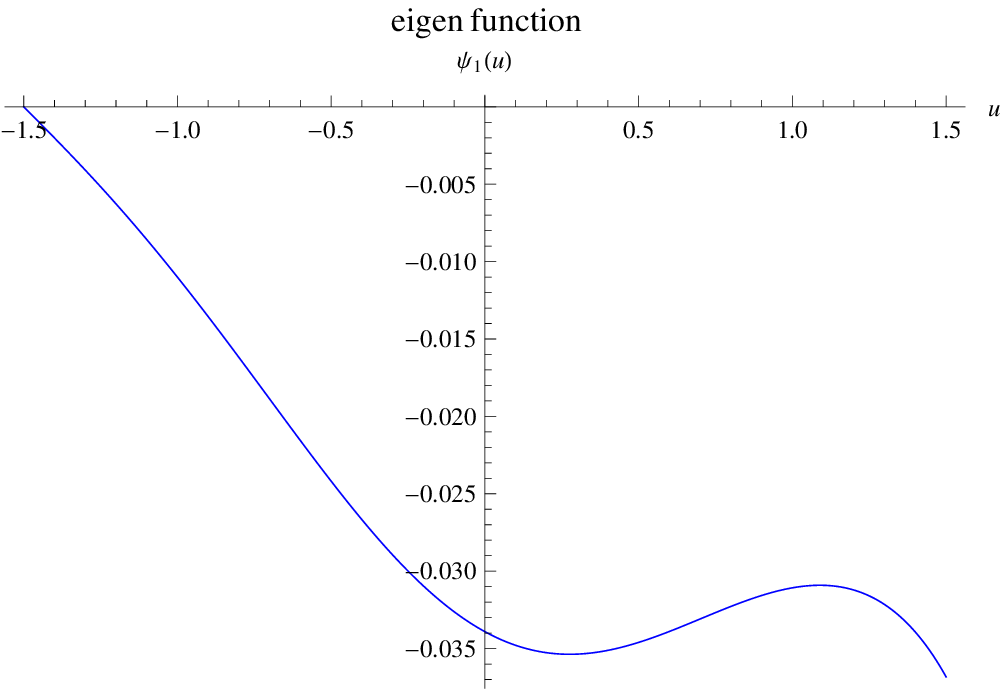}
\label{fig:subfig2}}
\subfigure[$\lambda_{1}=-0.5$]
{\includegraphics[scale=0.4]{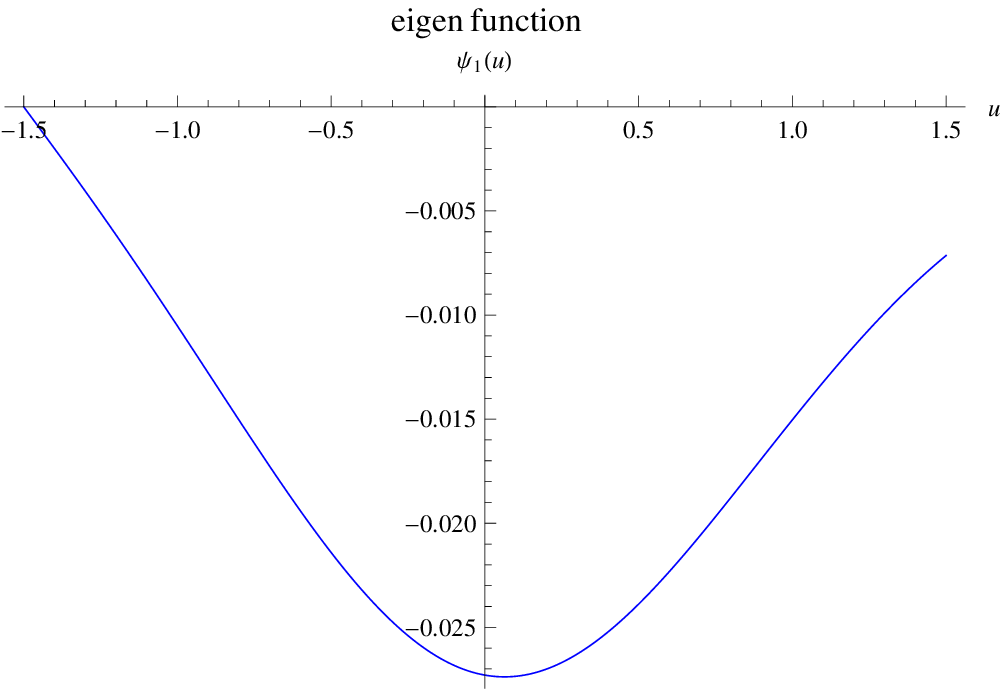}
\label{fig:subfig3}}
\subfigure[$\lambda_{1}=-0.4$]
{\includegraphics[scale=0.4]{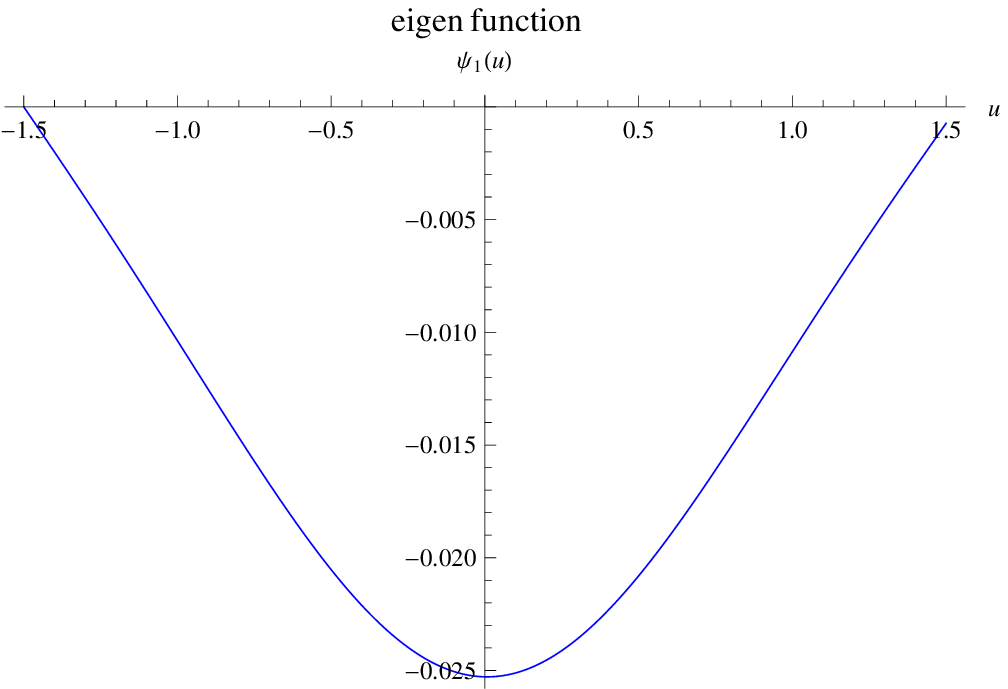}
\label{fig:subfig4}}
\subfigure[$\lambda_{1}=-0.3$ ]
{\includegraphics[scale=0.4]{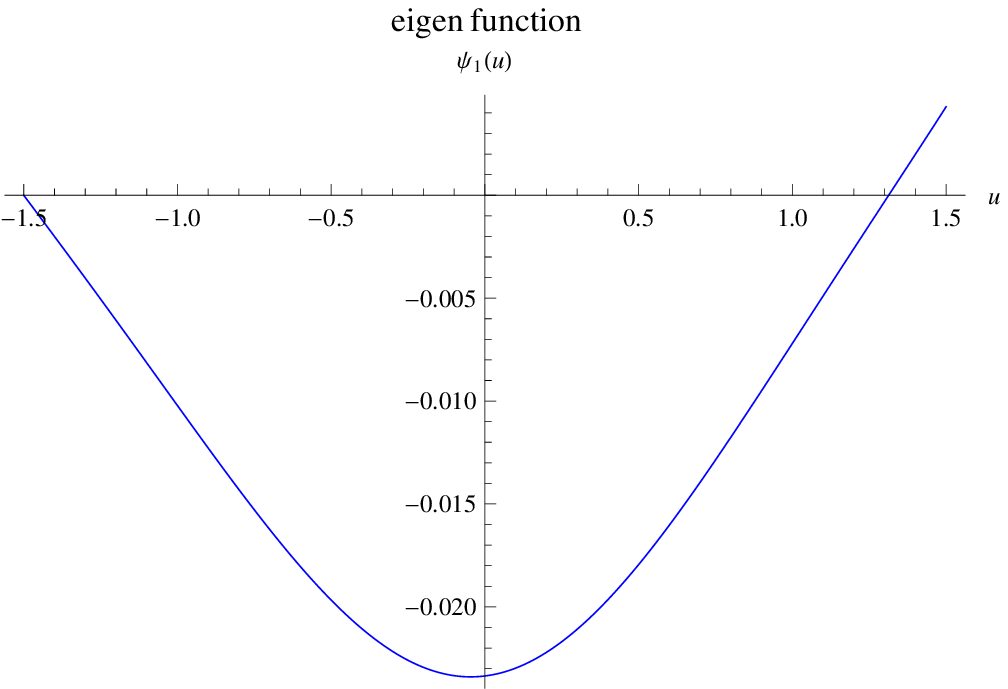}
\label{fig:subfig5}}
\subfigure[$\lambda_{1}=-0.385$ {\bf (correct)}]
{\includegraphics[scale=0.4]{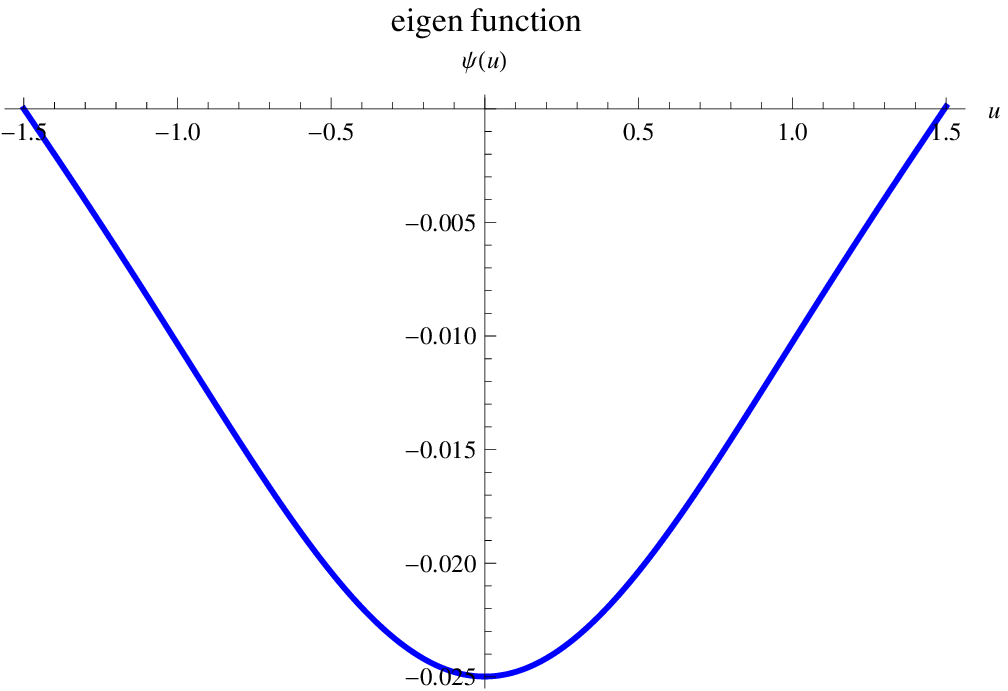}
\label{fig:subfig6}}
\caption{Plot of $\psi_{1}(u)$ for $m=0$,$u_0=1.5$ taking different $\lambda_{1}$, in order to find the correct eigenvalue. We start using the value that we get from AIM, which is (a). (f) is plot of the correct eigenfunction and therefore the correct eigenvalue is $-0.385$ . }
\label{fig:eigenfunction}
\end{figure}

To illustrate this, let us take the example of $u_0=1.5 ,m=0$. 
The plot of the eigenfunction using the eigenvalue $\lambda _1 = -0.66$ (
which we got using AIM) and assuming $(\frac{d\psi_1}{du})_{u=-1.5} = -0.02$ 
has been shown in Fig.~\ref{fig:subfig1}. 
Fig.~\ref{fig:subfig6} shows the correct eigenfunction obtained in the 
above way. The correct eigenvalue is $-0.385$ .


\begin{references}
\bibitem{lagrange} J. L. Lagrange, Oeuvres, Vol. 1 (1760).
\bibitem{plateau}  J. Plateau, {\em  Recherches exp´rimentales et th´orique sur les ﬁgures d’´quilibre d’une masse
liquide sans pesanteur}, M´m Acad Roy Belgiuque {\bf 29} (1849).
\bibitem{isenberg} C. Isenberg, {\em The science of soap films and soap
bubbles}, Dover Publications, N.Y., USA (1992).
\bibitem{osserman} R. Osserman, {\em A survey of minimal surfaces}, Dover
Publications, N. Y., USA (1986).
\bibitem{durand} L. Durand, {\em Stability and oscillations of a soap film: An analytic treatment}, Am. J. Phys, {\bf 49}, 334 (1981).
\bibitem{ito} M. Ito and T. Sato, {\em In situ observation of a soap-film catenoid- a simple educational physics experiment }, Eur.J.Phys.{\bf{31}},357-365 (2010).
\bibitem{benamar} M. Ben Amar, P.P. da Silva, N. Limodin, A. Langlois, 
M. Brazovskaia, C. Even, I.V. Chikina and P. Pieranski, {\em Stability and
vibrations of catenoid-shaped smectic films}, Eur. Phys. J. {\bf B 3}, 
197 (1998).
\bibitem{goldstein} H. Goldstein, C. Poole and J. Safko, {\em Classical Mechanics}, Pearson Education, India (2009) 
\bibitem{beeson} Michael Beeson, {\em Notes on Minimal Surfaces},\\
\url{http://www.michaelbeeson.com/research/papers/IntroMinimal.pdf}
\bibitem{kuiper} Lecture notes by H.Kuiper (Arizona State University), {\em Sturm-Liouville problems}, \url{http://math.la.asu.edu/~kuiper/462files/SturmLiouville.pdf}
\bibitem{dennery}P. Dennery and A. Krzywicki, {\em Mathematics for Physicists}, Dover Publications, New York, USA (1996).
\bibitem{ciftci} H. Ciftci, R. L. Hall and N. Saad, {\em Asymptotic iteration method for eigenvalue problems},  J. Phys. {\bf A} Math. Gen. {\bf 36} 11807.
\bibitem{cho } H. T. Cho, A. S. Cornell, J. Doukas,and Wade Naylor, {\em Black hole quasinormal modes using the asymptotic iteration method }, Class.Quant.Grav. {\bf 27},155004 (2010). 
\end{references}
\end{document}